\documentclass[aps,prl,reprint,groupedaddress]{revtex4-1}

\usepackage{amsmath}
\usepackage{graphicx}

\begin{document}


\title{A unified framework for information integration based on information geometry}


\author{Masafumi Oizumi$^{1,2}$}
\email{oizumi@brain.riken.jp}
\author{Naotsugu Tsuchiya$^{2,3,4}$}
\author{Shun-ichi Amari$^1$}
\affiliation{
$^{1}$ RIKEN Brain Science Institute, Wako, Saitama 351-0198, Japan
\\
$^{2}$ School of Psychological Sciences, Faculty of Biomedical and Psychological Sciences, Monash University, Clayton, Victoria 3800, Australia
\\
$^{3}$ Monash Institute of Cognitive and Clinical Neuroscience, Monash University, Clayton, Victoria 3800, Australia
\\
$^{4}$ Advanced Telecommunications Research Institute International, Keihanna Science City, Kyoto 619-0288, Japan
}


\date{\today}

\begin{abstract}
We propose a unified theoretical framework for quantifying spatio-temporal interactions in a stochastic dynamical system based on information geometry. In the proposed framework, the degree of interactions is quantified by the divergence between the actual probability distribution of the system and a constrained probability distribution where the interactions of interest are disconnected. This framework provides novel geometric interpretations of various information theoretic measures of interactions, such as mutual information, transfer entropy, and stochastic interaction in terms of how interactions are disconnected. The framework therefore provides an intuitive understanding of the relationships between the various quantities. By extending the concept of transfer entropy, we propose a novel measure of integrated information which measures causal interactions between parts of a system. Integrated information quantifies the extent to which the whole is more than the sum of the parts and can be potentially used as a biological measure of the levels of consciousness.
\end{abstract}


\maketitle

\section{Introduction}
There have been many attempts to quantify interactions between elements in a stochastic system. Information theory has played a pivotal role in this goal, leading to various measures of interactions such as multi-information \cite{Mcgill1954,Watanabe1960,Amari2001}, transfer entropy \cite{Schreiber2000}, stochastic interaction \cite{Ay2001,Ay2015}, and integrated information \cite{Tononi2004, Tononi2008}. In this study, we propose a unified theoretical framework for quantifying spatio-temporal interactions in a stochastic dynamical system based on information geometry \cite{Amari2007}, which is useful for elucidating various measures and their relations. In particular, we focus on the concept of integrated information and its related measures \cite{Ay2001,Barrett2011,Ay2015}. Utilizing the unified framework, we propose a novel measure of integrated information, called `geometric integrated information' $\Phi_G$. 

The concept of integrated information was proposed by Tononi in Integrated Information Theory (IIT) \cite{Tononi2004,Tononi2008,Oizumi2014}. Integrated information is designed for measuring the degree of causal interactions between parts of a system and the amount of information integrated in a system. IIT postulates that levels of consciousness correspond to the amount of integrated information in a system \cite{Tononi2004}, and this has been supported by experiments conducted in various types of loss of consciousness \cite{Massimini2005,Ferrarelli2010,Rosanova2012,Casali2013}. 

\begin{figure}[t!]
\begin{center}
\centerline{\includegraphics[width=0.3\textwidth]{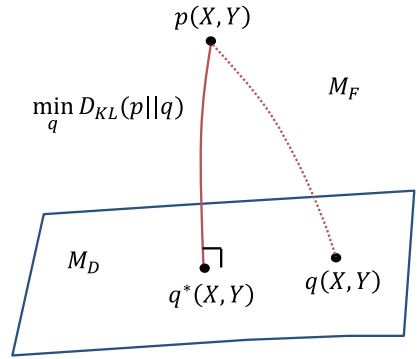}}
\caption{Information geometric picture for minimizing the KL divergence between the full model $p(X,Y)$ and disconnected model $q(X,Y)$. \label{fig:proj}} 
\end{center}
\end{figure}

Consider a joint probability distribution $p(X,Y)$ of the past states $X$ and the present states $Y$ of a system, where information about $X$ is integrated and transmitted to $Y$. Further, consider another probability distribution $q(X,Y)$, whose degree of freedom is constrained in terms of the way information is transmitted from $X$ to $Y$. More precisely, in $q(X,Y)$, some elements of $X$ are disconnected from other elements of $Y$, prohibiting information transmission over the branches that connect them. We call the former $p(X,Y)$ a `full model' and the latter $q(X,Y)$ a `disconnected model'. We propose to quantify the degree of interactions using the Kullback-Leibler (KL) divergence from the full to the disconnected model based on information geometry. A set of the disconnected model form a submanifold, $\mathcal{M}_D$, inside the entire manifold of the full model, $\mathcal{M}_F$. Given $p(X,Y)$, we search for the closest point $q^*(X,Y)$ to $p(X,Y)$ within the submanifold $\mathcal{M}_D$ (Fig. \ref{fig:proj}). The closest point is the minimizer of the KL divergence between $p(X,Y)$ and $q(X,Y)$ and is found by projecting $p(X,Y)$ to the submanifold. The minimized KL divergence can then be interpreted as `information loss' caused by disconnections of information transmission branches. 

This framework provides unified interpretations of various measures of interactions (Fig. \ref{fig:KL}) and offers a way to quantify these measures in a hierarchical manner which clarifies the relationships between them. We show that mutual information between $X$ and $Y$ corresponds to information loss when all the time-lagged interactions between $X$ and $Y$ are broken (Fig. \ref{fig:KL}(a)) and that transfer entropy from one element $x_i$ to another element $y_j$ corresponds to information loss when the transmission branch from $x_i$ to $y_j$ is eliminated (Fig. \ref{fig:KL}(b)). By extending transfer entropy, we propose a new measure of integrated information, which quantifies all causal interactions between parts of a system Fig. \ref{fig:KL}(c)). By construction, it is easy to see that the proposed measure of integrated information naturally satisfies an important theoretical requirement: Integrated information is non-negative and not more than the mutual information between $X$ and $Y$. A previously proposed measure of integrated information, called stochastic interaction \cite{Ay2001,Barrett2010,Ay2015}, does not satisfy this because stochastic interaction quantifies not only time-lagged interactions but also equal-time interactions in the present states $Y$ (Fig. \ref{fig:KL}(d)).

\section{Spatio-temporal interactions}

Consider a stochastic dynamical system in which the past and present states of the system are given by $X=(x_1,x_2,\cdots,x_N)$ and $Y=(y_1,y_2,\cdots,y_N)$, respectively. The spatio-temporal interactions of the system are characterized by the joint probability distribution $p(X,Y)$. In a dynamical system $p(X,Y)$, there are two types of interactions. Interactions between elements at the same time are called equal-time interactions. They can be quantified by analyzing only $p(X)$ or $p(Y)$. Interactions across different time points are called time-lagged interactions. They can be quantified from the conditional probability distribution $p(Y|X)$. Time-lagged interactions are also known as `causal' interactions \cite{Granger1988,Schreiber2000}, in a sense of `causality' that is statistically inferred from a limited amount of observation, which does not necessarily mean actual physical causality. Here, we use the term `causality' in this context and focus on quantifying causal interactions.

\subsection{Mutual information: Total causal interactions}
\begin{figure}[t!]
\begin{center}
\centerline{\includegraphics[width=0.5\textwidth]{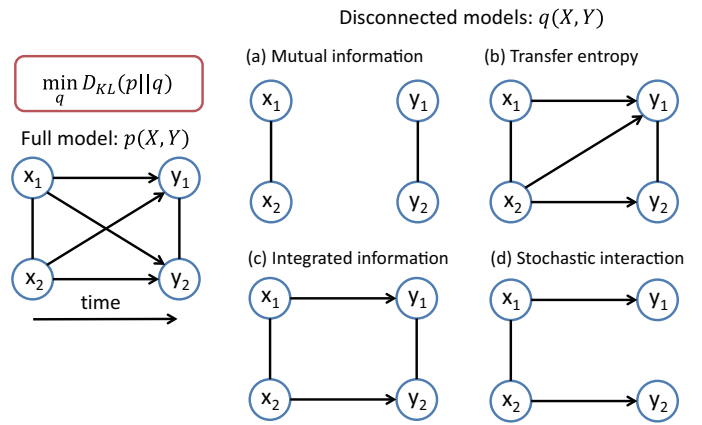}}
\caption{Minimizing the KL divergence between the full and the disconnected model (a)-(d) lead to various information theoretic quantities; (a) Mutual information, (b) Transfer entropy, (c) Integrated information, and (d) Stochastic interaction. Constraints imposed on the disconnected model are graphically shown. \label{fig:KL}} 
\end{center}
\end{figure}

First, we quantify the total degree of causal interactions between the past and the present states. We break the interactions between $X$ and $Y$ by forcing $X$ and $Y$ to be independent. Consider a manifold of probability distributions $\mathcal{M}_F$, where each point in the manifold represents a probability distribution $p(X,Y)$ (a full model) specified by $2N$ variables. Consider also a manifold $\mathcal{M}_I$ where $X$ and $Y$ are independent and there are no causal interactions between $X$ and $Y$. A probability distribution $q(X,Y)$ (a disconnected model), which is represented as a point in the manifold $\mathcal{M}_I$, is expressed as 
\begin{equation}
q(X,Y) = q(X)q(Y), \label{eq:MI_cons}
\end{equation}
and it is graphically represented in Fig. \ref{fig:KL}(a). The actual probability distribution $p(X,Y)$ is represented as a point outside the submanifold $\mathcal{M}_I$ (Fig. \ref{fig:proj}). In information geometry, dissimilarity between the two probability distributions is quantified by KL divergence. 
\begin{equation}
D_{KL}(p||q) = \int dX dY p(X,Y) \log \frac{p(X,Y)}{q(X,Y)}.
\end{equation}
We consider finding the best approximation of $p(X,Y)$ with $q(X,Y)$, which corresponds to minimizing the KL divergence between $p(X,Y)$ and $q(X,Y) \in \mathcal{M}_I$ (Fig. \ref{fig:proj}). The KL divergence is minimized by orthogonally projecting the point $p(X,Y)$ in the manifold $\mathcal{M}_I$ according to the projection theorem in information geometry \cite{Amari2007} (Fig. \ref{fig:proj}). The KL divergence is minimized when the marginal distributions of $q(X,Y)$ over $X$ and $Y$ are both equal to those of the actual distribution $p(X,Y)$, i.e., $q(X)=p(X)$ and $q(Y)=p(Y)$. The minimized KL divergence is given by
\begin{align}
\min_{q} D_{KL}(p||q) &= H(Y) - H(Y|X), \\
&= I(X;Y)
\end{align}
where $H$ is the entropy and $I(X;Y)$ is the mutual information between $X$ and $Y$. Thus, we can interpret the mutual information between $X$ and $Y$ as the total causal interactions between $X$ and $Y$. 

\subsection{Transfer entropy: Partial causal interactions}
Next, we quantify a partial causal interaction from one element to another in the system. We partially break a causal interaction from $x_i$ to $y_j$ by imposing the constraint that $x_i$ and $y_j$ are conditionally independent given the past states of the other variables except for $x_i$. We denote the set of the past states that excludes only $x_i$ by $\tilde{X}_i$. With this notation, the constraint imposed on the disconnected model $q(X,Y)$ is expressed as (Fig. \ref{fig:KL}(b))
\begin{equation}
q(y_j|X) = q(y_j|\tilde{X}_i). \label{eq:TE_cons_general}
\end{equation}
The KL divergence is minimized when $q(X)=p(X)$, $q(y_j|X)=p(y_j|\tilde{X}_i)$, and $q(\tilde{Y}_j|X,y_j)=p(\tilde{Y}_j|X,y_j)$. The minimized KL divergence is equal to the conditional transfer entropy from $x_i$ to $y_j$ given all the other past states $\tilde{X}_i$ except for the $i$-th element;
\begin{align}
\min_{q} D_{KL}(p||q) &= H(y_j|\tilde{X}_i) - H(y_j|X), \label{eq:general} \\ 
&= TE(x_i \to y_j | \tilde{X}_i), 
\end{align}
where $TE(x_i \to y_j | \tilde{X}_i)$ is the conditional transfer entropy from $x_i$ to $y_j$ given $\tilde{X}_i$. Thus, we can interpret the conditional transfer entropy as the degree of the partial causal interaction from $x_i$ to $y_j$.

\subsection{Integrated information}
Finally, we derive a new measure of integrated information by extending the concept of transfer entropy. Integrated information quantifies the degree of all causal interactions between parts of the system. Consider partitioning a system into $m$ parts, $P_1$, $P_2$, ..., $P_m$. We break all causal interactions between the different parts of the system by imposing the following constraints (Fig. \ref{fig:KL}(c));
\begin{equation}
q(Y[P_i]|X) = q(Y[P_i]|X[P_i]) \text{ ($\forall i$)}, \label{eq:phi_cons}
\end{equation}
where $X[P_i]$ and $Y[P_i]$ represent the past and the present states of the elements in the $i$-th subsystem $P_i$. To quantify integrated information, we consider a manifold $\mathcal{M}_G$ constrained by Eq. \ref{eq:phi_cons}. Note that within $\mathcal{M}_G$, the present states in a part $P_i$ only depend on the past states of itself and thus, the transfer entropies from one part $P_i$ to another part $P_j$ ($i \neq j$) are all 0. Now, we propose a new measure of integrated information, called `geometric integrated information' $\Phi_G$ as the minimized KL divergence between the actual distribution $p(X,Y)$ and the disconnected distribution $q(X,Y)$ within $\mathcal{M}_G$;
\begin{equation}
\Phi_G = \min_{q} D_{KL}(p||q).
\end{equation}
The manifold $\mathcal{M}_G$ formed by the constraints for integrated information (Eq. \ref{eq:phi_cons}) includes the manifold $\mathcal{M}_I$ formed by the constraints for mutual information (Eq. \ref{eq:MI_cons}), i.e., $\mathcal{M}_I \subset \mathcal{M}_G$. Since minimizing the KL divergence in a larger space always leads to a smaller value, $\Phi_G$ is always smaller than or equal to the mutual information $I(X;Y)$;
\begin{equation}
0 \leq \Phi_G \leq I(X;Y).
\end{equation}
The key concept of the integrated information is that it should quantify the extent to which the whole is more than the sum of parts \cite{Tononi2004,Ay2015}. Thus, it should be non-negative and upper bounded by the total amount of information in the whole system $I(X;Y)$ \cite{Oizumi2015}. $\Phi_G$ naturally satisfies this important theoretical requirement because it is derived from the minimization of the KL divergence. 

Importantly, $\Phi_G$ is not equal to the sum of the conditional transfer entropies of the deleted interactions. Moreover, the sum of the conditional transfer entropies can exceed the mutual information between $X$ and $Y$, which is the total amount of causal interactions in the whole system. As a simple illustrative example, consider a system consisting of two binary units, each of which takes one of the two states, $0$ or $1$. Assume that the probability distribution of the past states of $x_1$ and $x_2$ is a uniform distribution, i.e., $p(x_1,x_2) = 1/4$ and the present states of $y_1$ and $y_2$ are always equal, $y_1=y_2$, and are determined by the XOR gate of the past states $x_1$ and $x_2$. In this case, the transfer entropies from $x_1$ to $y_2$ or $x_2$ to $y_1$ are both 1, i.e., $TE(x_1 \to y_2 | x_2)= TE(x_2 \to y_1 | x_1) = 1$. However, the mutual information between $X$ and $Y$ is also 1, $I(X;Y)=1$ and thus, the sum of the transfer entropies is larger than the mutual information; $TE(x_1 \to y_2 | x_2) +  TE(x_2 \to y_1 | x_1) > I(X;Y)$. 

The sum of the conditional Granger causality from one element to another has been proposed as a measure of integration, termed `causal density' \cite{Seth2011}. The sum of the conditional Granger causality is equivalent to the sum of the conditional transfer entropies for Gaussian variables \cite{Barnett2009}. As explained above, the sum of conditional transfer entropies, or equivalently causal density, can exceed the mutual information between $X$ and $Y$. Thus, it does not satisfy the important theoretical requirement as an appropriate measure of integrated information. 

\subsection{Stochastic interactions}
Another measure, called stochastic interaction \cite{Ay2001,Ay2015}, was proposed as a different measure of integrated information \cite{Barrett2011}. In the derivation of stochastic interaction, Ay considered a manifold $\mathcal{M}_S$ where the conditional probability distribution of $Y$ given $X$ is decomposed into the product of the conditional probability distributions of each part (Fig. \ref{fig:KL}(d)) \cite{Ay2001,Ay2015};
\begin{equation}
q(Y|X) = \prod^m_{i=1} q(Y[P_i]|X[P_i]). \label{eq:SI_cons}
\end{equation}
This constraint satisfies that for the integrated information in Eq. \ref{eq:phi_cons}. Thus, $\mathcal{M}_S \subset \mathcal{M}_G$. In addition to that, this constraint further satisfies conditional independence among the present states of parts, $Y[P_i]$, given the past states in the whole system $X$;
\begin{equation}
q(Y|X) = \prod^m_{i=1} q(Y[P_i]|X). \label{eq:equal_cons}
\end{equation}
This constraint corresponds to breaking equal-time interactions among the present states of the parts given the past states of the whole, whereas the constraint in Eq. \ref{eq:phi_cons} corresponds to breaking only time-lagged interactions between the present and past states (Fig. \ref{fig:KL}(d)).

The KL divergence is minimized when $q(X)=p(X)$ and $q(Y[P_i]|X[P_i])=p(Y[P_i]|X[P_i])$. The minimized KL divergence is equal to stochastic interaction $SI(X;Y)$:
\begin{align}
\min_{q} D_{KL}(p||q) &= \sum_i H(Y[P_i]|X[P_i]) -H(Y|X), \\
&= SI(X;Y).
\end{align}
In contrast to the manifold $\mathcal{M}_G$ considered for $\Phi_G$, the manifold $\mathcal{M}_S$ formed by the constraints for stochastic interaction (Eq. \ref{eq:SI_cons}) does not include the manifold $\mathcal{M}_I$ formed by the constraints for the mutual information between $X$ and $Y$ (Eq. \ref{eq:MI_cons}). This is because not only causal interactions but also equal-time interactions are broken in $\mathcal{M}_S$. Stochastic interaction can therefore exceed the total degree of causal interactions in the whole system, which is not suitable as a measure of integrated information \cite{Oizumi2015}.

\section{Relation to Granger causality}
\begin{figure}[t!]
\begin{center}
\centerline{\includegraphics[width=0.4\textwidth]{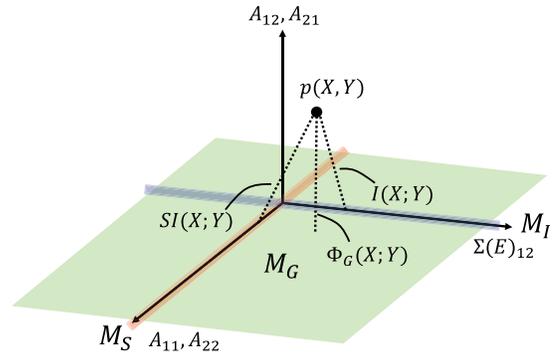}}
\caption{Relationships between manifolds for mutual information $\mathcal{M}_I$ (the gray line), stochastic interaction $\mathcal{M}_S$ (the orange line), and integrated information $\mathcal{M}_G$ (the green plane) in the Gaussian case. $\mathcal{M}_I$ is the line where $A=0$, $\mathcal{M}_S$ is the line where $\Sigma(E)_{12}$ and $A_{12},A_{21}$ are 0, and $\mathcal{M}_G$ is the plane where $A_{12},A_{21}$ are 0. \label{fig:gaussian}} 
\end{center}
\end{figure}

In this section, we show a close relationship between the proposed measure of integrated information $\Phi_G$ and multivariate Granger causality by computing $\Phi_G$ when the probability distribution of a system $p(X,Y)$ is Gaussian. Consider a following multivariate autoregressive model,
\begin{equation}
Y = AX + E, \label{eq:AR_full}
\end{equation}
where $X$ and $Y$ are the past and present states of a system, $A$ is the connectivity matrix, and $E$ is Gaussian random variables, which are uncorrelated over time. The multivariate autoregressive model is the generative model of a multivariate Gaussian distribution. Regarding Eq. \ref{eq:AR_full} as a full model, we consider the following as a disconnected model: 
\begin{equation}
Y = A'X + E'.
\end{equation}
The constraints for $\Phi_G$ (Eq. \ref{eq:phi_cons}) correspond to setting the off-diagonal elements of $A'$ to 0:
\begin{equation}
A'_{ij} = 0 (i \neq j).
\end{equation}
It is instructive to consider the constraints for the other information theoretic quantities introduced above; the constraints for (a) mutual information, (b) transfer entropy from $x_1$ to $y_2$, and (d) stochastic interaction. They correspond to (a) $A'=0$, (b) $A'_{21}=0$ and (d) the off-diagonal elements of $A'$ and $\Sigma(E)'$ being 0, respectively. Fig. \ref{fig:gaussian} shows the relationship between the manifolds formed by the constraints for mutual information $\mathcal{M}_I$, stochastic interaction $\mathcal{M}_S$, and integrated information $\mathcal{M}_G$. We can see that $\mathcal{M}_I$ and $\mathcal{M}_S$ are included in $\mathcal{M}_G$. Thus, $\Phi_G$ is smaller than $I(X;Y)$ or $SI(X;Y)$. On the other hand, there is no inclusion relation between $\mathcal{M}_I$ and $\mathcal{M}_S$.

By differentiating the KL divergence between the full model $p(X,Y)$ and the disconnected model $q(X,Y)$ with respect to $\Sigma(X)'^{-1}$ (the covariance of $X$ in the disconnected model), $A'$, and $\Sigma(E)'^{-1}$, we can find the minimum of the KL divergence using the following equations:
\begin{align}
\Sigma(X)' = \Sigma(X), \label{eq:W} \\
(\Sigma(X)(A-A')\Sigma(E)'^{-1}))_{ii} =0, \\
\Sigma(E)' =\Sigma(E)+(A-A')\Sigma(X)(A-A')^T. \label{eq:V}
\end{align}
By substituting Eqs. \ref{eq:W}-\ref{eq:V} into the KL divergence, we obtain
\begin{equation}
\Phi_G = \frac{1}{2} \log \frac{|\Sigma(E)'|}{|\Sigma(E)|}. \label{eq:phi_D}
\end{equation}
$|\Sigma(E)|$ is called the generalized variance, which is used as a measure of goodness of fit, i.e., the degree of prediction error, in multivariate Granger causality analysis \cite{Geweke1982,Barrett2010}. In the Gaussian case, $\Phi_G$ can be interpreted as the difference in the prediction error on comparison of the full and disconnected model, in which the off-diagonal elements of $A'$ are set to 0. Thus, $\Phi_G$ is consistent with multivariate Granger causality analysis based on the generalized variance. $\Phi_G$ can be rewritten as the difference between the conditional entropy in the full model and that in the disconnected model, 
\begin{equation}
\Phi_G = H(q(Y|X)) - H(p(Y|X)). \label{eq:phi_DH}
\end{equation}

For comparison, mutual information, transfer entropy, and stochastic interaction are given as follows: 
\begin{align}
I(X;Y) &= \frac{1}{2} \log  \frac{|\Sigma(X)|}{|\Sigma(E)|}, \\
TE(x_i \to y_j | x_j) &= \frac{1}{2} \log \frac{\Sigma(E)^*_{jj}}{\Sigma(E)_{jj}}, \\
SI(X;Y) &= \frac{1}{2} \log \frac{\Sigma(E)^*_{11}\Sigma(E)^*_{22}}{|\Sigma(E)|}, 
\end{align}
where $\Sigma(E)^*_{jj}$ ($j=1,2$) is the covariance of the conditional probability distribution $p(y_j|x_j)$. 

\section{Hierarchical structure}
\begin{figure}[t!]
\begin{center}
\centerline{\includegraphics[width=0.4\textwidth]{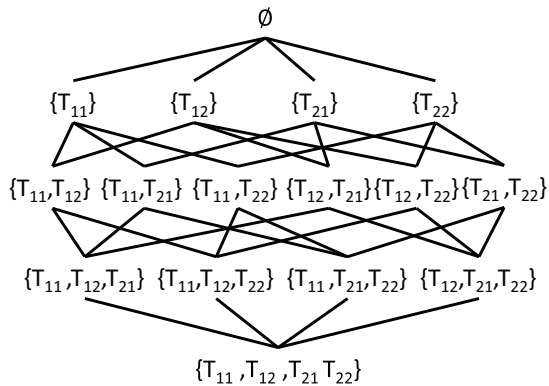}}
\caption{A hierarchical structure of the disconnected models where time-lagged interactions are broken in a system consisting of two units. All possible combinations of interactions retained in the disconnected model are displayed. If two models are related with the addition or removal of one interaction, they are connected by a line. The KL divergence between the full and the disconnected model increases from bottom to top. \label{fig:hierarchy}} 
\end{center}
\end{figure}

We can construct a hierarchical structure of the disconnected models and then use it to systematically quantify all possible combinations of causal interactions \cite{Ay2011}. For example, in a system consisting of two elements, there are four time-lagged interactions, $x_1 \to y_1$, $x_1 \to y_2$, $x_2 \to y_1$ and $x_2 \to y_2$, which are denoted by $T_{11}$, $T_{12}$, $T_{21}$ and $T_{22}$, respectively. While we consider only the cross interactions, $T_{12}$ and $T_{21}$ for transfer entropy and integrated information, we can also quantify self-interactions $T_{11}$ and $T_{22}$ by imposing the corresponding constraints, such as $q(y_1|x_1,x_2)=q(y_1|x_2)$ and $q(y_2|x_1,x_2)=q(y_2|x_1)$, respectively. A set of all possible disconnected models forms a partially ordered set with respect to KL divergence between the full and the disconnected models (Fig. \ref{fig:hierarchy}). If one disconnected model is related to another one with the removal or inclusion of interactions, there is an ordering relationship in the KL divergence. Such related models are connected by a line in Fig. \ref{fig:hierarchy}. From the bottom to the top, information loss increases as more interactions are broken. Note that there is no ordering relationship between the disconnected models at the same level of the hierarchy. At the top, all four interactions are destroyed, and thus information loss is maximized, which corresponds to the mutual information $I(X;Y)$. The hierarchical structure generalizes all related measures mentioned in this article and provides a clear perspective on the relationship among different measures. 

\section{Conclusion}
In this paper, we propose a unified framework for quantifying spatio-temporal interactions in a stochastic dynamical system based on information geometry. By evaluating KL divergence between the full and disconnected models, we provided unified interpretations of various measures of interactions, including transfer entropy. By extending transfer entropy in this framework, we derived a novel measure of integrated information $\Phi_G$. We showed that $\Phi_G$ naturally satisfies the important theoretical requirement as a proper measure of integrated information: Integrated information is less than or equal to the mutual information in the whole system, i.e., $0 \leq \Phi_G \leq I(X;Y)$ \cite{Oizumi2015}. By analytically calculating $\Phi_G$ in the Gaussian case, we revealed a close relationship between $\Phi_G$ and multivariate Granger causality based on the generalized variance. Finally, we pointed out that the hierarchical structure of the disconnected models enables us to quantify any combination of spatio-temporal interactions in a systematic manner from a unified perspective. We expect that the proposed framework can be utilized in diverse research fields, including neuroscience \cite{Deco2015,Boly2015} where network connectivity analysis has been an active research topic \cite{Bullmore2009}. We also expect that our measure of integrated information will be particularly useful for consciousness researchers \cite{Lee2009,Chang2012,King2013} because information integration is considered to be a key prerequisite of conscious information processing in the brain \cite{Tononi2004}.

\begin{acknowledgments}
M.O. was supported by a Grant-in-Aid for Young Scientists (B) from the Ministry of Education, Culture, Sports, Science, and Technology of Japan (26870860). N.T. was supported by Precursory Research for Embryonic Science and Technology from the Japan Science and Technology Agency (3630), the Future Fellowship (FT120100619) and the Discovery Project (DP130100194) from the Australian Research Council.
\end{acknowledgments}

\bibliography{OizumiTsuchiyaAmari_arXiv.bib}

\end{document}